\input{aipcheck}

\documentclass[
   ,final            
  ]
  {aipproc}

\layoutstyle{6x9}

\begin{document}

\title{Particle acceleration in electron-ion jets}

\vspace*{-.4cm}
\author{K.-I. Nishikawa}{
address={National Space Science and Technology Center,
  Huntsville, AL 35805}
}

\author{P. Hardee}{
address={Department of Physics and Astronomy,
  The University of Alabama,
  Tuscaloosa, AL 35487}
}

\author{C. B. Hededal}{
address={Niels Bohr Institute, Department of Astrophysics, Juliane
Maries Vej 30, 2100 København \/{O}, Denmark}}


\author{G. Richardson}{
address={Department of Mechanical and Aerospace Engineering
University of Alabama in Huntsville Huntsville, AL 35899}}

\author{R. Preece}{
address={Department of Physics,
  University of Alabama in Huntsville,
  Huntsville, AL 35899 and National Space Science and Technology Center,
  Huntsville, AL 35805}}

\author{H. Sol}{
address={LUTH, Observatore de Paris-Meudon, 5 place Jules Jansen
92195
   Meudon Cedex, France}}

\author{G. J. Fishman}{
address={NASA-Marshall Space Flight Center, \\
National Space Science and Technology Center,
  Huntsville, AL 35805}}
\author{C. Kouvelioutou}{
  address={NASA-Marshall Space Flight Center, \\
National Space Science and Technology Center,
  Huntsville, AL 35805}
}
\author{Y. Mizuno}{
  address={National Space Science and Technology Center,
  Huntsville, AL 35805} 
}

\vspace*{-1.0cm}
\begin{abstract}
Weibel instability created in collisionless shocks is responsible
for particle (electron, positron, and ion) acceleration. Using a 3-D
relativistic electromagnetic particle (REMP) code, we have
investigated particle acceleration associated with a relativistic
electron-ion  jet fronts propagating into an ambient plasma  without
initial magnetic fields with a longer simulation system in order to
investigate nonlinear stage of the Weibel instability and its
acceleration mechanism. The current channels generated by the Weibel
instability induce the radial electric fields. The $z$ component of
the Poynting vector (${\bf E} \times {\bf B}$) become positive in
the large region along the jet propagation direction. This leads to
the acceleration of jet electrons along the jet. In particular the
$E \times B$ drift with the large scale current channel generated by
the ion Weibel instability accelerate electrons effectively in both
parallel and perpendicular directions.
\end{abstract}

\keywords{Weibel instability, RPIC simulations, particle
acceleration} \classification{98.54.Cm,98.62.Nx,98.70.Qy,01.30.Cc}


\maketitle


\vspace*{-.5cm}
\section{Simulation Results}

\vspace*{-0.3cm}
 Recent PIC simulations using injected relativistic
electron-ion jets show that acceleration occurs within the
downstream jet, rather than by the scattering of particles back and
forth across the shock as in Fermi acceleration [1-5, 8].

In this report we present new simulation results of particle
acceleration and magnetic field generation for relativistic jet
shocks using 3-D relativistic electromagnetic particle-in-cell
(REMP) simulations. Our new simulations with further diagnostics has
revealed the mechanism of particle acceleration with the Weibel
instability.

In the collisionless shock generated behind the head of the
relativistic jet the Weibel instability is excited in the downstream
region.  The instability generates current filaments elongated along
the streaming direction and associated transverse loop magnetic
fields. These current channels generate the induced radial electric
fields which are mainly perpendicular to the loop transverse
magnetic fields. Since the Weibel instability is convective, the the
current channels propagate with the jets. The ${\bf E} \times {\bf
B}$ also propagate with the jet. Some particles are trapped and
accelerated.

 Figure 1 (left) shows the $z$ component of ${\bf E}
\times {\bf B}$ with the $x$ and $y$ components of magnetic field
(by the arrows) generated by the ion Weibel instability in the $x -
y$ plane at $z/\Delta = 430$ for the electron-ion jet injected into
an unmagnetized ambient electron-ion plasma at $t = 59.8/\omega_{\rm
pe}$. At the region with large positive $z$ component electrons are
accelerated along the jet propagation direction. Consequently, at
the front part of jet (one third) electrons are accelerated more
than $10 \gamma v$ as shown in the left panel [6].

\begin{figure}[ht]
  \resizebox{13pc}{!}{\includegraphics{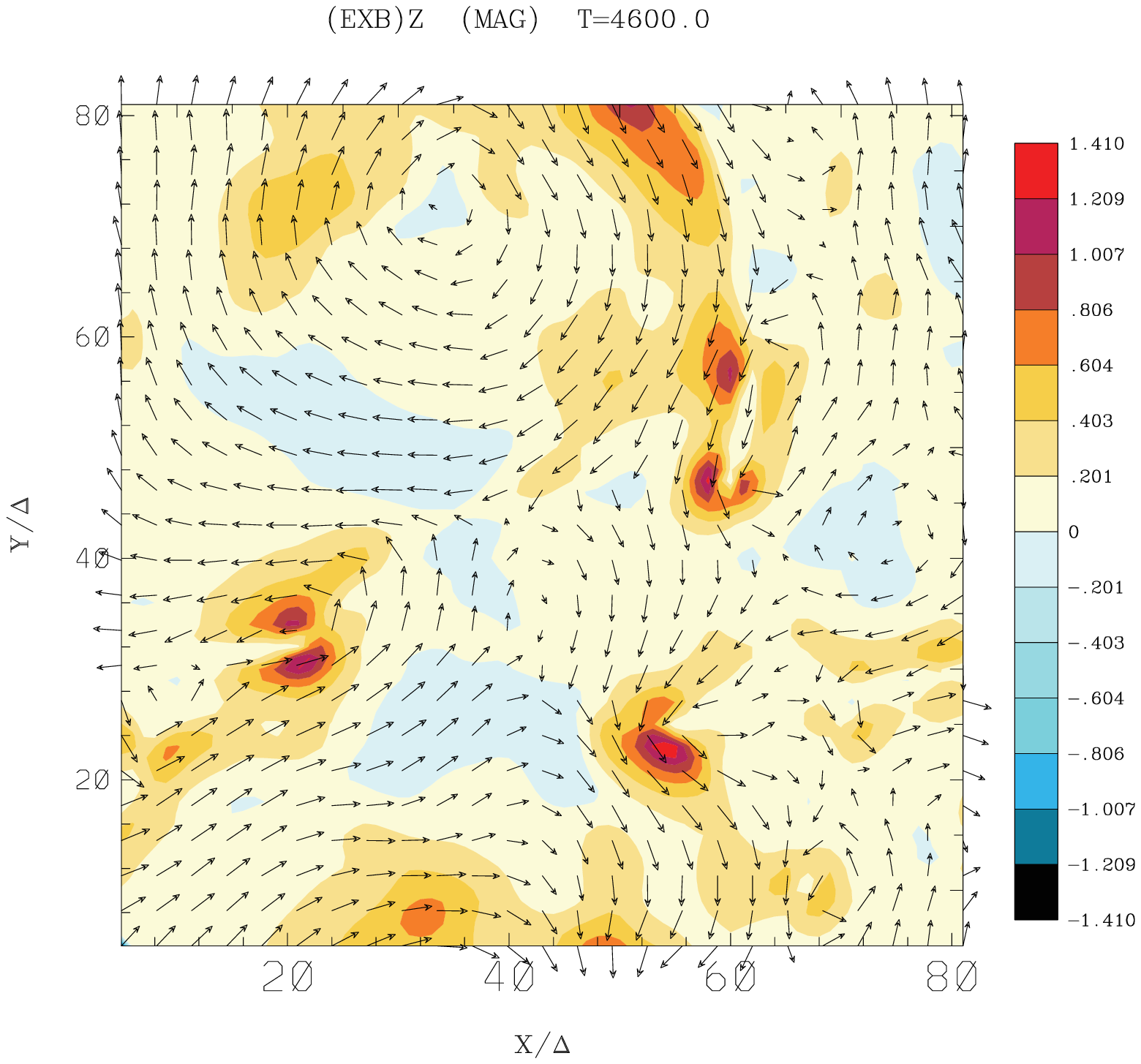}}
  \resizebox{16pc}{!}{\includegraphics{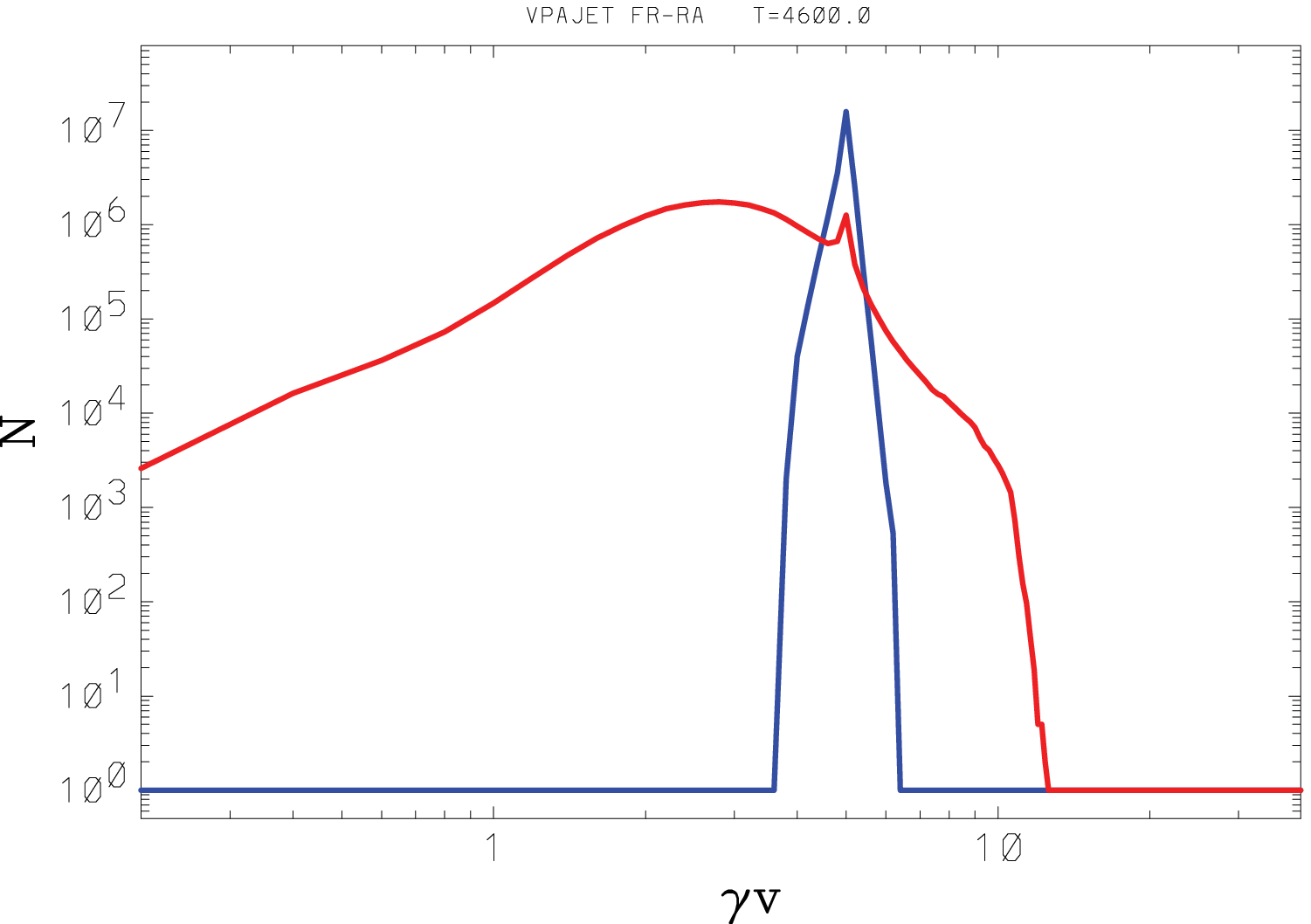}}
\caption{Left and right panels show the $z$ component of ${\bf E}
\times {\bf B}$ at $z/\Delta = 430$ and velocity distributions of
jet electrons (One third of jet electrons in the front region is
plotted by red curves and on third of rear part of jet is plotted by
blue curves)  at $t = 59.8/\omega_{\rm pe}$, respectively. Jet
electrons are binned as a function of $\gamma v$, where $\gamma = (1
-(v^{2}_{\rm x} +v^{2}_{\rm y} +v^{2}_{\rm z})/c^{2})^{-1/2}$.}
\end{figure}

The $z$ component of ${\bf E} \times {\bf B}$ becomes mainly
positive on $z-x$ plane (not shown). This component accelerate jet
electrons
effectively as also shown in Fig. 1a in [2]. 
Certainly, the negative component decelerate particles. This $E
\times B$ drift acceleration leads to sufficient acceleration in
parallel and perpendicular directions.

We have also performed simulations with widely distributed Lorentz
factor of jet electrons and positrons which are assumed to be
created by
the photon annihilation [7]. 



\vspace*{-0.2cm}
\begin{theacknowledgments}

\vspace*{-0.2cm}
  This research (K.N.) is
partially supported by the National Science Foundation awards ATM
9730230, ATM-9870072, ATM-0100997, INT-9981508, and AST-0506719.
The simulations have been performed on IBM p690 (Copper) at the
National Center for Supercomputing Applications (NCSA) which is
supported by the National Science Foundation.

\end{theacknowledgments}


\bibliographystyle{aipproc}   



\end{document}